\documentclass{pasa}%

\usepackage{upgreek}
\usepackage{graphicx}
\usepackage{subfigure}
\usepackage{hyperref}

\title[PyGEDM]{A comparison of Galactic electron density models using PyGEDM}

\author[D. C. Price]{
D.C. Price$^{1,2,3,}$\thanks{email: danny.price@curtin.edu.au},
C. Flynn$^2$,
A. Deller$^2$

\affil{$^1$International Centre for Radio Astronomy Research, Curtin University, Bentley WA 6102, Australia}
\affil{$^2$Centre for Astrophysics and Supercomputing,
Swinburne University of Technology,
Hawthorn VIC 3122 Australia}
\affil{$^3$Department  of  Astronomy, 
University  of  California  Berkeley, 
Berkeley  CA  94720}
}%

\newcommand{\udm}[0]{$\rm{pc\,cm}^{-3}$}  

\jid{PASA}
\doi{10.1017/pas.\the\year.xxx}
\jyear{\the\year}

\usepackage{aas_macros}
\usepackage{hyperref} 
\hypersetup{colorlinks,citecolor=blue,linkcolor=blue,urlcolor=blue}

\hypersetup{draft}

\begin{document}

\begin{frontmatter}
\maketitle

\begin{abstract}
Galactic electron density distribution models are crucial tools for estimating the impact of the ionised interstellar medium on the impulsive signals from radio pulsars and fast radio bursts.  The two prevailing Galactic electron density models are YMW16 \citep{Yao:2017} and NE2001 \citep{Cordes:2002}. Here, we introduce a software package \textsc{PyGEDM} which provides a unified application programming interface (API) for these models and the YT20 \citep{Yamasaki:2019} model of the Galactic halo. We use \textsc{PyGEDM} to compute all-sky maps of Galactic dispersion measure (DM) for YMW16 and NE2001, and compare the large-scale differences between the two. In general, YMW16 predicts higher DM values toward the Galactic anticentre. YMW16 predicts higher DMs at low Galactic latitudes, but NE2001 predicts higher DMs in most other directions. We identify lines of sight for which the models are most discrepant, using pulsars with independent distance measurements. YMW16 performs better on average than NE2001, but both models show significant outliers. We suggest that future campaigns to determine pulsar distances should focus on targets where the models show large discrepancies, so future models can use those measurements to better estimate distances along those line of sight. We also suggest that the Galactic halo should be considered as a component in future GEDMs, to avoid overestimating the Galactic DM contribution for extragalactic sources such as FRBs.

\end{abstract}

\begin{keywords}
pulsars:general --- stars:distances --- ISM:structure --- fast radio bursts
\end{keywords}
\end{frontmatter}

\section{Introduction}

Electron density models---models of the distribution of free electrons in the Galaxy---are routinely used to convert a dispersion measure (DM) along a given line-of-sight to an estimated distance, and vice versa. More broadly, they are used extensively in studies of Galactic composition, to describe scintillation and interstellar scattering, and to differentiate between extragalactic fast radio bursts (FRBs) and giant pulse emission from Galactic pulsars.

The DM is related to the integral free electron number density $n_e$ between Earth and a source at distance $d$. A frequency-dependent time delay, $\Delta t$, is imparted to electromagnetic radiation travelling through the ionised plasma along this sight line. For two observing frequencies, $\nu_1$ and $\nu_2$, the time delay is related to DM by:
\begin{equation}
	{\rm{DM}} \triangleq \int_{0}^{d} n_e\,dl = K \left(\frac{1}{\nu_{1}^{2}}-\frac{1}{\nu^{2}_2}\right)^{-1}\Delta t,
\end{equation}
where in SI units, $K$ is given by
\begin{equation}
K=\frac{2\pi m_e c}{e^2} = 241.0331786(66)\, \rm{GHz^{-2}\,cm^{-3}\,pc\,s^{-1}}.
\end{equation}
Here, $e$ is electron charge, $m_e$ is electron mass, and $c$ is the speed of light\footnote{Several approximations for $K$ are commonplace, and DM is also weakly sensitive to electron temperature and the presence of other, heavier charged particles; see the extensive discussion in \citet{Kulkarni:2020}.}.

The impulsive, broad-band nature of radio pulsar emission means that observations of pulsars can provide measurements of DM---and thus the number of electrons along their sight line. If the electron density along the sight line is known, one can determine the distance to the pulsar, but this requires a Galactic electron density model (GEDM). 


Over the years, GEDMs have been derived from independent distance measurements to pulsars (see \S 2). The prevailing GEDMs are NE2001  \citep{Cordes:2002, Cordes:2003} and the Yao-Machester-Wang model \citep[YMW16,][]{Yao:2017}. While YMW16 benefits from more recent data, and was shown in \citet{Yao:2017} to give improved pulsar distance estimates, the NE2001 model is still in widespread use. 


Further comparison of the two models was conducted in  \citet{Deller:2019} on a sample of 57 pulsars with VLBI parallax measurements, which indicated 1) the the YWM16 model performs `somewhat better' on high-latitude pulsars in the sample, 2) both models show large errors for some objects, and 3) both models tend to underestimate distances for the sample pulsars.

There is growing evidence from low-DM Fast Radio Bursts (FRBs) that the models may overestimate the Galactic DM contribution---despite neither model including a DM contribution arising from the Galactic halo. The \citet{CHIME:2019} report the repeating FRB180916.J0158+65 to have a DM of 349.2(3) \udm, very close to the 330 \udm\, Galactic contribution as reported by YMW16. As FRB180916.J0158+65 has now been localized to a nearby spiral galaxy \citep{Marcote:2020}, it is clear that YMW16  overestimates DM along this line-of-sight.  Similarly, FRB 180430 is placed within the Galaxy by YMW16, but not NE2001 \citep{Qiu:2019}.

Clearly, there is room for improvement. Simple improvements to the models can be made by refitting parameters after including new independent pulsar distance measurements along individual sightlines, such as the large sample in \citet{Deller:2019}. Complementary to this approach---and the subject of this paper---qualitative analysis of how and where the models differ can help identify limitations within the models, and how these may be improved. 

Improved estimates of Galactic DM contribution are of particular concern for FRB studies, as inaccurate estimates may limit the use of FRBs for cosmology. \citet{Kumar:2019} argues that for FRBs to be useful as a distance measure, bias in the noncosmological contributions to DM must be kept to under 0.6\% of the total DM value: a challenging requirement. 

Following \citet{Yamasaki:2019}, the total DM for an extragalactic source can be expressed as a sum of four components:
\begin{equation}
	\rm{DM}_{\rm{obs}} = \rm{DM}_{\rm{ISM}} + \rm{DM}_{\rm{halo}} + \rm{DM}_{\rm{IGM}} + \rm{DM}_{\rm{host}}
\end{equation}
where $\rm{DM}_{\rm{ISM}}$ is the contribution from the interstellar medium (ISM) in
the Milky Way (MW) disk, $\rm{DM}_{\rm{halo}}$ is the contribution from the extended Galactic
halo, $\rm{DM}_{\rm{IGM}}$ is that from the intergalactic medium (IGM), and $\rm{DM}_{\rm{host}}$
is that from the host galaxy (including the source's local environment). Note that on some sight lines, DM contributions from intervening galaxy halos may also need to be considered \citep[e.g.][]{Prochaska:2019b, Connor:2020}. Only the $\rm{DM}_{\rm{IGM}}$ component---recently measured by \citet{Macquart:2020}---is of interest for inferring cosmological distance; systematically overestimating or underestimating $\rm{DM}_{\rm{ISM}}$ by using an inaccurate GEDM may thus bias or confound cosmological efforts.


In this paper, we present comparisons of the two models, made using a new Python package called \textsc{PyGEDM} that provides a unified interface to the YMW16 and NE2001 codes. Previous comparisons \citep[e.g.,][]{Yao:2017, Deller:2019} have focused on how well the models predict the DM of pulsars with independent distance measurements; here, we compare their estimates on large angular scales. 
 
 This paper is organized as follows. In \S2, we provide a brief overview of NE2001 and YMW16, and introduce the empirical relation between DM, $N_H$, and $A_{\rm{V}}$. In \S3, we introduce the \textsc{PyGEDM} package. In \S4, comparisons of YMW16 and NE2001 are made, and model performance is compared against recent pulsar and FRB measurements. The paper concludes with a discussion of model limitations and recommendations for future measurements and improvements.

\section{Galactic Electron Density Models}

\begin{figure*}
\centering 
\includegraphics[width=2.0\columnwidth]{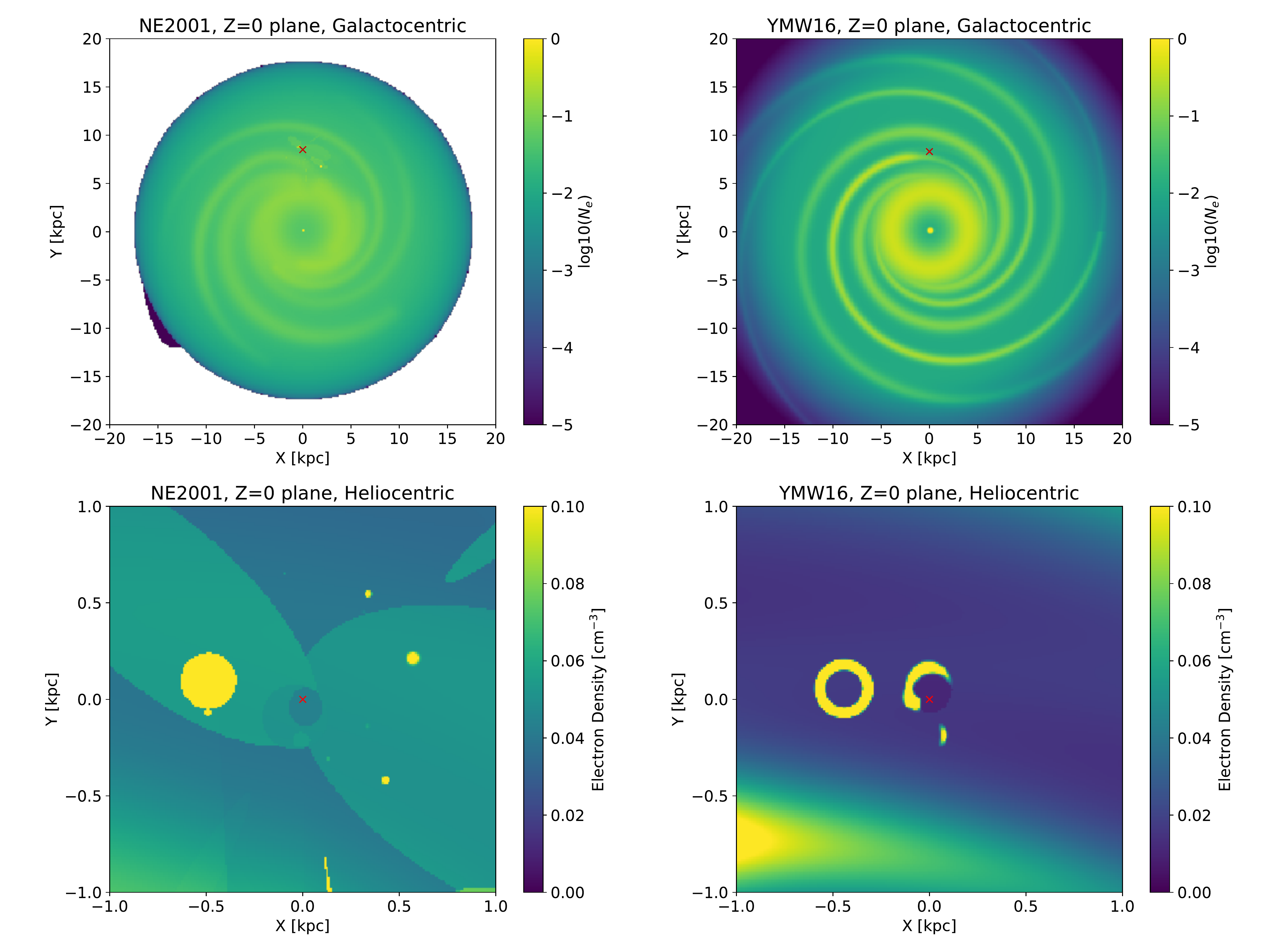}%
\caption{Electron density of NE2001 (left) and YMW16 (right) models, in the Galactic plane ($z=0$).  The  NE2001 model extends to $\pm$17\,kpc, whereas YMW16 extends to a radius $\pm$30\,kpc. The Sun (red cross) is placed at $x=0$, $y=8500$\,pc, $z=0$ in NE2001, and at $x=0$, $y=8300$\,pc, $z=6$\,pc in YMW16. The top panels show large-scale Galactic structure; differences in the spiral arm structure are visible. The bottom panels show the local ISM in a $\pm$1\,kpc region centred about the Sun. The large ellipses in NE2001 (bottom left) correspond to a `local superbubble' and `low density region', which are not included in the YMW16 model. The local `clumps' of NE2001, also not used in YMW16, are also visible as small circular regions. The local ISM in the YMW16 model (bottom right) has visibly fewer components; identifiable are the Gum nebula, local bubble, Loop I, and Carina-Saggitarius spiral arm. }
\label{fig:galaxy-ne}. 
\end{figure*}
 
Forming a GEDM requires a set of distance measurements to pulsars paired with  DM measurements; see \citet{Verbiest:2012} for an overview. The signature of annual parallax is the most common model-independent distance measurement, and can be obtained using very long baseline interferometry (VLBI) imaging, or by fitting for variations in the arrival time of radio pulses in precision pulsar timing solutions. For pulsars within the Galactic disk, kinematic distance measurements can be made by use of 21-cm absorption spectra to convert radial velocities of HI clouds along the sight line to distances (however, this requires a Galactic rotation model). A third method is to use an association with nebulae, a globular cluster, or an optical counterpart.

Two of the earliest models of Galactic electron density were the LMT85 model \citep{Manchester:1981, Lyne:1985} and a model by \citet{Vivekanand:1982}, henceforth VK82. The LMT85 model was based upon kinematic distance measurements from HI absorption for 36 pulsars, and consisted only of three components: a thin disk, a term dependent upon height above the Galactic plane, and a model for the Gum Nebula. The VK82 model was similarly simple, but was derived independently using data from the Second Molonglo pulsar survey \citep{Manchester:1978}. 

These models were superseded by the TC93 model \citep{Cordes:1991, Taylor:1993}, formed using 74 independent pulsar distance measurements. Measurements of interstellar scattering were also included the model. As our understanding of Galactic structure advanced, and the number of independent distance measurements to pulsars increased, it became clear that an updated model was required to fix shortcomings in TC93.

\subsection{NE2001}
NE2001, as detailed in \citet{Cordes:2002, Cordes:2003}, and plotted in the lefthand panels of Fig.\,\ref{fig:galaxy-ne}, addressed many of the TC93 model's shortcomings. NE2001 was formed from 112 pulsar distance measurements (i.e. independent of the GEDM), and 269 scattering measurements. Broadly, NE2001 assumes smoothly varying, large-scale components, then adds in perturbations by small-scale underdense or overdense regions. The model places the Galactic centre at a distance of 8.5 kpc from the Sun, and uses a right-handed Galactocentric coordinate system ($x$, $y$, $z$) with the $x$-axis parallel to $l$=90$^\circ$, and $y$-axis aligned toward $l$=180$^\circ$. The Sun is located in the disk, at ($x$=0, $y$=8500 pc, $z$=0). The Galactic structure consists of a thin Gaussian annulus and thick axisymmetric disk, and spiral arms. Local components---the hot `local bubble' surrounding the Sun, Gum Nebula, Vela Supernova Remnant (SNR), Loop 1, and a few other features---are also included in the model, along with a Galactic centre component. NE2001 also invokes so-called `clumps' and `voids', to account for sightlines where measurements suggest over- and under-dense regions respectively.

NE2001 uses an an iterative approach to parameter fitting \citep[see \S5 of][]{Cordes:2003}. Preliminary values from the TC93 were used, then parameters for large-scale components were fit by use of a likelihood function, followed by parameters from the local ISM; this process was then iterated.
  
\subsection{YMW16} 
YMW16, as detailed in \citet{Yao:2017}, and plotted in the righthand panels of Fig.\,\ref{fig:galaxy-ne}, has the significant advantage of 15 years of additional data to use when fitting their model. Over the intervening years, systematic issues with NE2001 were identified, which also informed the YMW16 model. Firstly, NE2001 systematically underestimates the $z$-distance for pulsars at high galactic latitude \citep{Lorimer:2006}, due partly to the scale height for the thick disk being too small, which is evidenced by new observational measurements \citep{Gaensler:2008, Savage:2009}. The NE2001 model was also found to overpredict distances for some local pulsars \citep{Chatterjee:2009}. 

 YMW16 uses 189 independent pulsar distance measurements, but unlike NE2001 does not make use of interstellar scattering measurements, arguing that  scattering is generally dominated by a few regions along the path to a pulsar, so scattering measurements do not inform about large-scale structure.  
 
In YMW16, the Sun is placed at ($x$=0, $y$=8300 pc $z$=6 pc), which includes an offset from the Galactic plane as reported in \citet{Joshi:2016}.
Like NE2001, YMW16 uses three major Galactic components.  The first, an axisymmetric thick disk, is modelled in a similar fashion to NE2001.  However, its thin disk component is modelled both radially and vertically with $sech^2$ functions, and it uses a 4-spiral-arm model \citet{Hou:2014} in contrast to the modified spiral pattern used in NE2001.
Seven local features---the Local Bubble, regions of enhanced density nearby the Local Bubble, the Gum Nebula, Loop I  \citep{Berkhuijsen:1971}, a enhanced region in the Carina arm, and a low density pocket in the Sagittarius tangential region---are included in the YMW16 model, similar to NE2001. YMW16 also provides a models for the Magellanic clouds and IGM. Beyond these features, YMW16 rejects the use of voids and clumps for pulsar-specific optimization, arguing this is poor practice as it leads to overfitting.  Further comparison of the features in the two models can be found in \S5.2 of \citet{Yao:2017}.

Overall, the YMW16 model has 117 parameters; 35 parameters are fitted by using an optimization routine, with the other parameters fixed to values from the literature. Parameter fitting was performed using the \textsc{PSwarm} `particle swarm' algorithm \citep[see \S4 of][]{Yao:2017}. 

\subsection{Galactic halo models}

YMW16 and NE2001 do not model the DM contribution from the Galactic halo, $\rm{DM}_{halo}$. While modelling the halo is not necessary for pulsars within the Milky Way, a model of the halo is necessary for determining $\rm{DM_{host}}$ and $\rm{DM_{IGM}}$ extragalactic sources such as FRBs. Here, we provide a brief overview of estimates for $\rm{DM}_{halo}$, starting with a toy model where we assume baryons in the halo are distributed spherically and with uniform density out to the virial radius.

It is well established that the baryons residing in galaxies in the form of stars and cold or warm gas account for only a fraction of the baryons measured in the the $\Lambda$CDM model. These baryons are predominantly thought to reside in the IGM, but some are expected to reside within the virial radii of galaxies. The baryonic mass of the Milky Way that has been accounted for directly is $\approx 7 \times 10^{10}$ $M_\odot$ (Flynn et al 2006), while its dark matter halo has a mass order $1.2 \times 10^{12}$ $M_\odot$  and a virial radius of order 200 kpc (Wang et al 2015). Assuming the Milky Way has captured the cosmic fraction of baryons to dark matter ($\approx 0.15$), we estimate $\approx 1.3 \times 10^{11}$ hot baryons in the halo. The DM is then given by $R_{\mathrm{vir}} n_e$, where $n_e$ is the electron number density. Taking $M_{\rm B}$ is the mass in baryons in the halo, $m_p$ as the proton mass, and $R_{\rm{vir}}$ is the virial radius, the electron density $n_e$ is given by $(M_{\rm B}/m_p)/V$ where $V$ is $\frac{4}{3}\pi R_{\rm vir}^3$.
Inserting typical values yields DM $= 30$\,\udm\, for the Milky Way. 

This toy model is broadly consistent with what is found in hydrodynamical simulations of galaxy formation in a cosmological context. For example, \citet{Dolag:2015} analyse a Milky Way-like galaxy from such simulations, find a range of DM in the (lumpy) halo of 40 to 70\,\udm\, somewhat more than the simple estimate of 30\,\udm\ above. 

\subsubsection{Halo DM estimates via diffuse gas}

\citet{Prochaska:2019} (henceforth PZ19) estimate $\rm{DM}_{halo}$ by fitting two components: $T\sim10^4$\,K gas (which they refer to as ``cool'') and $T\sim10^6$\,K gas (referred to as ``hot''):
\begin{equation}
	\rm{DM}_{halo} = \rm{DM}_{halo,cool} + \rm{DM}_{halo,hot}.
\end{equation}
 For their cool component, they isolate high velocity clouds (HVCs) with velocities $>100$ kms$^{-1}$ from HI4PI data \citep{HI4PI:2016}, to produce a $N_{H,\rm{HVC}}$ map for HVCs only. These data are combined with column density measurements of Si$_{\rm{II}}$ and Si$_{\rm{III}}$ from the \citet{Richter:2017} HVC survey, which are dominant ions of silicon at $T\sim10^4$\,K. PZ19 finds an average value of $\rm{DM}_{halo,cool}\approx20$\,\udm. For the hot component, PZ19 analyzes  measurements of O$_{\rm{VI}}$ and O$_{\rm{VII}}$ X-ray absorption spectra  \citep{Fang:2015}, and estimate that the ionized plasma revealed by these tracers adds a contribution $\rm{DM}_{halo, hot}$=50--80\,\udm.

\citet{Das:2020} (henceforth D20) use a similar approach to PZ19, but do not directly differentiate between the disk and halo. Instead, they model the overall Galactic DM contribution, $\rm{DM}_{\rm{MW}}$, as a combination of four phases:
\begin{equation}
	\rm{DM}_{MW} = \rm{DM}_{cold} + \rm{DM}_{cool} + \rm{DM}_{warm} + \rm{DM}_{hot},
\end{equation}
where the four components refer to gas at $\sim$10$^4$K, $10^4$--$10^5$K, $10^5$--$10^{5.5}$K, and $>10^6$K, respectively.  Note that these designations differ from the temperature ranges typically used to refer to gas phases in the interstellar medium. Their ``cold'' and ``hot'' phases are found to be the primary contributors to $\rm{DM}_{\rm{MW}}$, by at least an order of magnitude. 

Combined, D20 find a median $\rm{DM}_{MW}=64^{+20}_{-23}$\,\udm, covering with a large scatter---two orders of magnitude, 33--172\,\udm\, (68\% CI). This scatter is not predicted by smooth disk + halo models, and is supportive of the hot halo component being inhomogeneous and anisotropic.

\citet{Keating:2020} use gas profile models of the Galactic halo and incorporate constraints from X-ray observations to compute predicted values of $\rm{DM_{halo}}$. Values of $\rm{DM}_{halo}$ below 55\,\udm\, are favoured; however predictions for models allowed by X-ray constraints span more than an order of magnitude, reaching as low as 6\,\udm.

\subsubsection{Halo DM estimates via pulsars and FRBs}

Another way to estimate the halo DM is to use pulsars at high galactic latitudes and in the Magellanic clouds. The clouds are sufficiently distant ($\sim 50$ kpc) that most of the DM in an NFW-like distribution of hot baryons is within their Galactocentric radii \citep{Navarro:1997}. Pulsars have a range of DMs in the Large Magellanic Cloud (LMC) from 65 to 200\,\udm\, (using data available in PSRCAT\footnote{http://www.atnf.csiro.au/research/pulsar/psrcat}; \citealp{manchesterHobbsetal2005}). Assuming that the lower DM value of 65\,\udm\, represents pulsars least affected by the LMC's own ISM, and that 50\,\udm\, of this is due to the disk ISM in the direction of the LMC (as estimated by both the NE2001 and YMW16 models), this yields a lower limit on the halo DM (in this direction) of approximately 15\,\udm. 

\citet{Platts:2020} have made a similar analysis using the DMs of Galactic pulsars combined with the DMs of published FRBs, introducing a kernel density estimation technique to find lower and upper bounds for $\rm{DM_{halo}}$. 
They place a constraint of  $-2<\rm{DM_{halo}}< 123$\,\udm (95\% C.I.), assuming a spherical distribution of the baryons. Tighter constraints may be derived using the \citet{Platts:2020} framework as the sample of FRBs grows, and in particular as more low DM (i.e. nearby) FRBs are found. 

A summary table of estimates for different model approaches is given in Tab.\,\ref{tab:halo-summary}.

\subsubsection{YT20}

Observational support that a simple symmetric spherical halo model is inadequate comes from X-ray observations of diffuse halo \citep[e.g.][]{Nakashima:2018}. They favour a two component model: a spherical component extending up to 200\,kpc, and a compact disk-like component that is geometrically distinct from the thick disk in ISM models.

\citet{Yamasaki:2019} (henceforth YT20) provide a two-component model fit to  observations of diffuse X-ray emission. YT20  predicts  $\rm{DM}_{halo}$ to be 30--245\,\udm \,over the whole sky, with a mean of 43\,\udm. 

\begin{figure}
\centering 
\includegraphics[width=1.0\columnwidth]{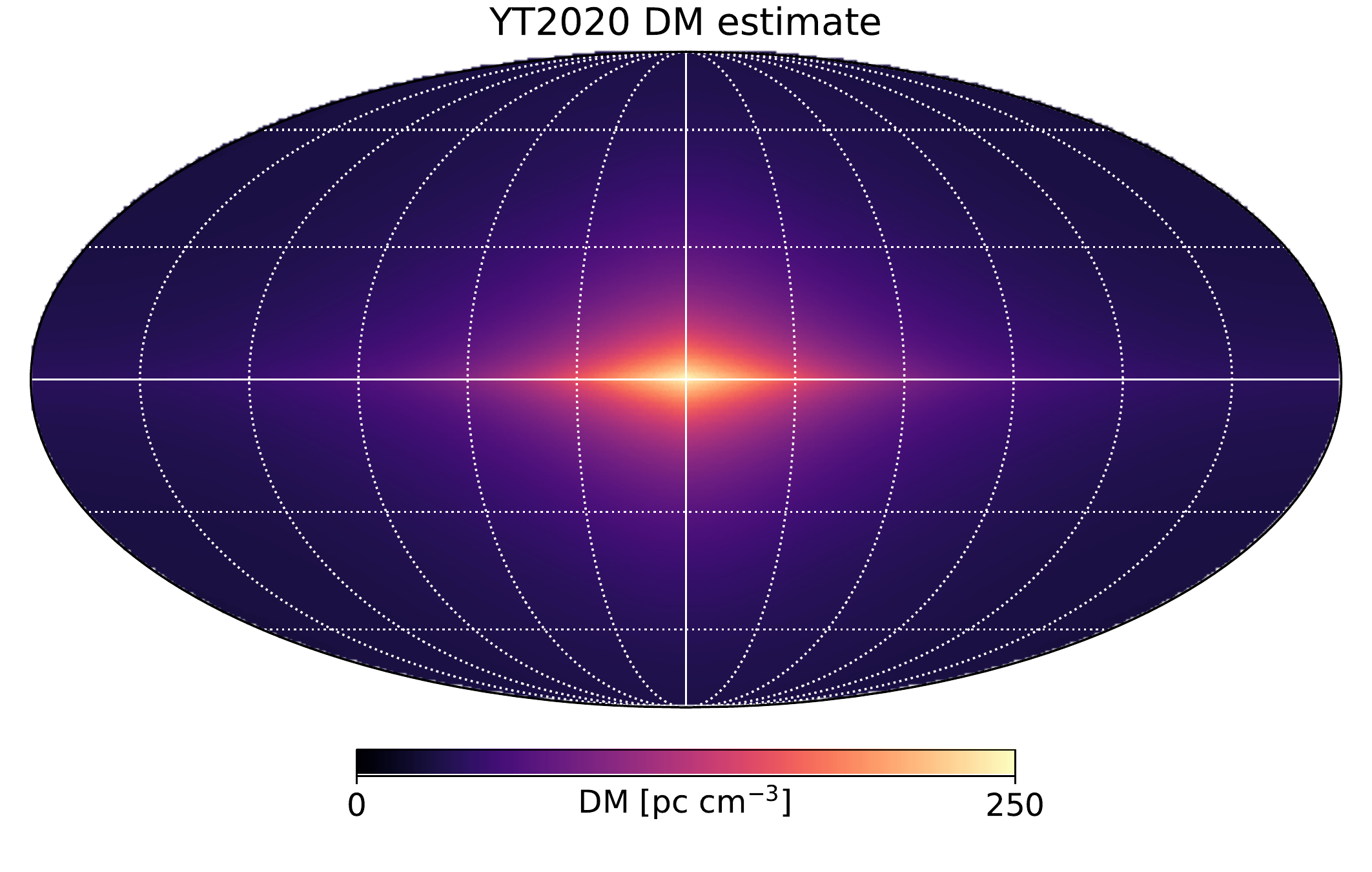}%
\caption{Estimates of $\rm{DM_{halo}}$ from the YT20 model \citep{Yamasaki:2019}. }
\label{fig:YT2020-dm-map}
\end{figure}

The YT20 model consists of a spherical halo that extends to the virial radius (200\,kpc), and a compact disk-like component. This disk-like component is differs in physical properties (i.e. temperature and geometrical shape) from the ISM's thick disk, as included in NE2001 and YMW16. 
An all-sky map of the YT20 model is shown in Fig.\,\ref{fig:YT2020-dm-map}. 

While the gas distribution of the halo disk-like component overlaps with the ISM thick disk component, the YT20 authors argue it is reasonable to add the DM prediction from YMW16/NE2001 to estimate the total Galactic DM budget:
\begin{equation}
	\rm{DM_{MW} = DM_{YT20} + DM_{YMW16}}.
\end{equation}
However, as the gas distribution of the ISM thick disk and YT20 disk-like component overlap, GEDM models may overestimate the density of the thick disk; adding YT20 and YW16 estimates of DM may result in an overestimate for $\rm{DM_{MW}}$.

Currently, YT20 is the only halo model with multiple components, giving rise to direction-dependent DM estimates. We provide an interface to YT20 as part of \textsc{PyGEDM}.

\begin{table}

\caption{\label{tab:halo-summary} Summary of halo DM contribution from model estimates. Note \citet{Das:2020} estimate is for the full Galactic DM contribution, $\rm{DM_{MW}}$.}
\centering
\begin{tabular}{lcc}
\hline
Halo model &  DM$_{\rm{halo}}$  &  DM$_{\rm{MW}}$ \\
       & (\udm)  & (\udm)   \\
\hline
\hline
Baryon count estimate  & 30      & -\\
LMC pulsar constraint  & $>$15   & -\\
\citet{Dolag:2015}     & 40--70  & -\\
\citet{Prochaska:2019} & 70--100 & -\\
\citet{Yamasaki:2019}  & 30--245 & -\\
\citet{Platts:2020}    & $-$2--123 & -\\
\citet{Keating:2020}   &  6--55 & -\\
\citet{Das:2020}       &  -  &  33--172 \\
\hline
\end{tabular}
\end{table}

\section{PyGEDM}

We have developed a Python package, \textsc{PyGEDM}, which provides access to the NE2001 and YMW16 models of the ISM, and YT20 model of Galactic halo. Unit and coordinate conversions are handled using the \textsc{Astropy} package \citep{Astropy:2013}, and \textsc{Healpy} \citep{Zonca:2019} is used to generate all-sky maps. The \textsc{PyGEDM} code is open source and freely available online\footnote{\url{https://github.com/FRBs/pygedm}}. 

\textsc{PyGEDM} can be installed via the Python package manager with a single command ({\tt pip install pygedm}), and build files are provided for the Docker\footnote{\url{https://www.docker.com}} containerization platform. A test suite is included to ensure code output is consistent with the YMW16 and NE2001 online interfaces. Searchable online documentation of the \textsc{PyGEDM} API is provided at \url{https://pygedm.readthedocs.io}.

\textsc{PyGEDM} provides a unified application programming interface (API) in Python, with the intention that it can be used as an upstream dependency for other projects. By doing so, changes to \textsc{PyGEDM}---for example, the addition of new GEDMs---can be immediately leveraged by downstream projects and data analysis codes. An example use case is for population synthesis codes such as \textsc{Frbpoppy}\footnote{\url{https://davidgardenier.github.io/frbpoppy/html/index.html}} \citep{Gardenier:2019} and \textsc{PsrPopPy}\footnote{\url{https://github.com/samb8s/PsrPopPy}} \citep{Bates:2014}, for FRB and pulsars respectively. These codes currently wrap a precompiled version of the NE2001 model. \textsc{PyGEDM} is used by  \textsc{Fruitbat}\footnote{\url{https://github.com/abatten/fruitbat}} \citep{Batten:2019}, which computes the redshift of FRBs for given cosmological models.

Both the YMW16 and NE2001 codes are written in performant, statically-compiled languages (C and Fortran, respectively). Python provides methods to interface with both C and Fortran code; in \textsc{PyGEDM} we use {\tt pybind11}\footnote{\url{https://pybind11.readthedocs.io/}} to interact with compiled library versions of the code. {\tt pybind11} is a library that exposes Python types in C++, and vice versa.

\subsection{An interface to NE2001 using {\tt f2c} and {\tt pybind11}}

Compiling the NE2001 code requires a Fortran compiler such as {\tt gfortran}. To provide an interface to NE2001 via {\tt pybind11}, and to avoid the need for a Fortran compiler, we used the {\tt f2c} Fortran to C conversion program to convert the NE2001 codebase to C.  While NE2001 is mostly compliant with the Fortran 77 standard, some reordering of statements was required to satisfy the {\tt f2c} utility. The converted C code is available within the \textsc{PyGEDM} repository.

Before deciding on using {\tt f2c} and {\tt pybind11}, we used the {\tt f2py} utility---part of the \textsc{Numpy} package \citep{vanderWalt:2011}---to generate compiled extension modules that can be used in Python. This required adding special comment lines defined by {\tt f2py} to the NE2001 Fortran code, which the Fortran compiler ignores but inform {\tt f2py} whether arguments are meant as inputs, outputs, or both. When the \textsc{PyGEDM} Python package is installed, a Fortran compiler is called to compile NE2001 code, then {\tt f2py} is run to compile Python-compatible shared objects. However, we found that this approach was not portable across different architectures and systems. As of writing, there is no native Fortran compiler for the Apple Silicon M1 architecture, and several users reported that installing \textsc{PyGEDM} via {\tt pip install pygedm} did not work. We also found issues setting up continuous integration (CI) testing with Fortran, and hence ultimately abandoned this approach.

Using {\tt f2c}-converted C code alleviates issues with Fortran compilers, but adds a requirement that the user has the {\tt f2c.h} header and corresponding library installed. We find {\tt pybind11} to be more robust and easier to debug during installation of the Python package. The {\tt pybind11} approach also allows a more Pythonic API: the C++ {\tt std::map} container is presented as a {\tt dict} to Python. We use this to return a dictionary of key-value pairs when the C++ function is called.  

Other open-source codes to allow Python access to NE2001 are available. \textsc{pyne2001}\footnote{\url{https://github.com/v-morello/pyne2001}} calls the NE2001 executable and parses the resulting text output; this approach requires no changes to the Fortran code, but is slower than access via {\tt pybind11}. The \textsc{ FRBs/ne2001}\footnote{\url{https://github.com/FRBs/ne2001}} code is a pure-Python re-implementation of the NE2001 code, not guaranteed to give identical results to the Fortran NE2001 code, and is considerably slower. The \textsc{FrbPoppy} and \textsc{PsrPoppy} codes compile NE2001 as a shared library, and then access it via Python {\tt ctypes}.

We tested the speed of these approaches by installing all software within a Docker container, and then calling the relevant Python API to calculate the DM for a point 10 kpc away in the direction of the Galatic centre ($b$=0, $l$=0). The Docker container was run on a Macbook Pro laptop (2020) with the Apple Silicon M1 chip, and the {\tt \%timeit} magic function was used in an iPython shell to find the average runtime. We found the raw {\tt ctypes} method to be the fastest at $\sim$110$\mu$s per call, with our {\tt pybind11} approach taking $\sim$790$\mu$s. We attribute the overhead to the construction of the {\tt std::map}, and conversion to \textsc{Astropy} quantities. {\tt pyne2001} is roughly 70 times slower than the {\tt ctypes} approach, at $\sim7.96$ms. Finally, the pure Python {\tt ne2001} implementation takes $\sim$2.6s: several orders of magnitude slower. 

Although speed is important in some use cases, code maintainability, adaptability, and usability are also important considerations. While our approach is slower than using raw {\tt ctypes}, overhead within Python, such as attribute lookup for variables within {\tt for} loops, is likely to be the main bottleneck for most programs using \textsc{PyGEDM}. An approach to speed up loops over multiple lines of sight would be to move the loop into C++, which could be achieved via the  built-in support for \textsc{Numpy} arrays in {\tt pybind11}.

\subsection{An interface to YMW16 and YT20}

As YMW16 is written in C, only minor changes to the underlying code were required in order to create Python bindings using {\tt pybind11}. We added a {\tt main.cpp} file, in which the {\tt pybind11} module is defined. As with the NE2001 interface, we modified the YMW16 API to return {\tt std::map} containers, which are presented in Python as dictionaries of key-value pairs.

The YT20 code in \textsc{PyGEDM} is adapted from {\tt DM\_halo\_yt2020\_numerical.py}, provided by S. Yamasaki. As there were no extra dependencies, integration with \textsc{PyGEDM} was straightforward; some minor changes to coding style were made for uniformity.

\subsection{The PyGEDM API}

The \textsc{PyGEDM} code provides the following methods:

\begin{itemize}
    \item {\tt dm\_to\_dist} -- Convert a DM to a distance for a given line-of-sight in Galactic ($l$, $b$) coordinates (NE2001 and YMW16).
    \item {\tt dist\_to\_dm} -- Convert a distance to a DM for a given line-of-sight (NE2001 and YMW16).
    \item {\tt calculate\_electron\_density\_xyz} -- Evaluate the electron density at a given Galactocentric ($x,y,z$) coordinate (NE2001 and YMW16).
    \item {\tt calculate\_electron\_density\_lbr} -- Evaluate electron density at a given distance along line-of-sight in Galactic coordinates ($l,b$), to a distance $r$ (NE2001 and YMW16).
    \item {\tt  generate\_healpix\_dm\_map} -- Generate an all-sky healpix map for a given distance $r$ (NE2001, YMW16 and YT20).
    \item {\tt calculate\_halo\_dm}  -- Compute $\rm{DM_{halo}}$ for a given ($l, b$) pointing (YT20).
    \item {\tt convert\_lbr\_to\_xyz} -- Convert Galactic ($l,b,r$) coordinates to Galactocentric ($x,y,z$) coordinates (NE2001 or YMW16 -- the two models place the Galactic centre at different distances).
\end{itemize}

Selection between YMW16, NE2001 and YT20 models is done by use of a {\tt method} argument. We envisage adding support for future models as they become available.

\subsection{Containerization}

Docker is an open-source containerization platform, which is used to package up code and its dependencies into a standardized executable that can be built and deployed in a reproducable way. As part of the \textsc{PyGEDM} repository, we provide a {\tt Dockerfile}, from which a Docker container for \textsc{PyGEDM} can be generated. 

\subsection{Integration testing}

Neither NE2001 or YMW16 are supplied with a testing framework, so we tested the output of \textsc{PyGEDM} against the online interfaces at \url{https://www.nrl.navy.mil/rsd/RORF/ne2001/} (NE2001, now defunct), and \url{https://www.atnf.csiro.au/research/pulsar/ymw16/} (YMW16). To ensure future development does not alter code output, we wrote unit tests using {\tt pytest} that check code output against known correct values. We have setup continuous integration (CI) testing using Github actions to automatically run the unit tests whenever code is pushed to the \textsc{PyGEDM} repository. 

Code coverage---a report of what parts of code are executed by the tests---is analysed using Codecov.io\footnote{\url{https://codecov.io}} for all Python code in \textsc{PyGEDM}; all Python code is covered by unit tests (i.e. 100\% code coverage). Full coverage ensures that all functions are tested for correctness. We have not attempted to write unit tests for the underlying NE2001 or YMW16 code, but suggest that future GEDMs should do so as a matter of course.

\subsection{PyGEDM web application}

\begin{figure}
\centering 
\includegraphics[width=1.0\columnwidth]{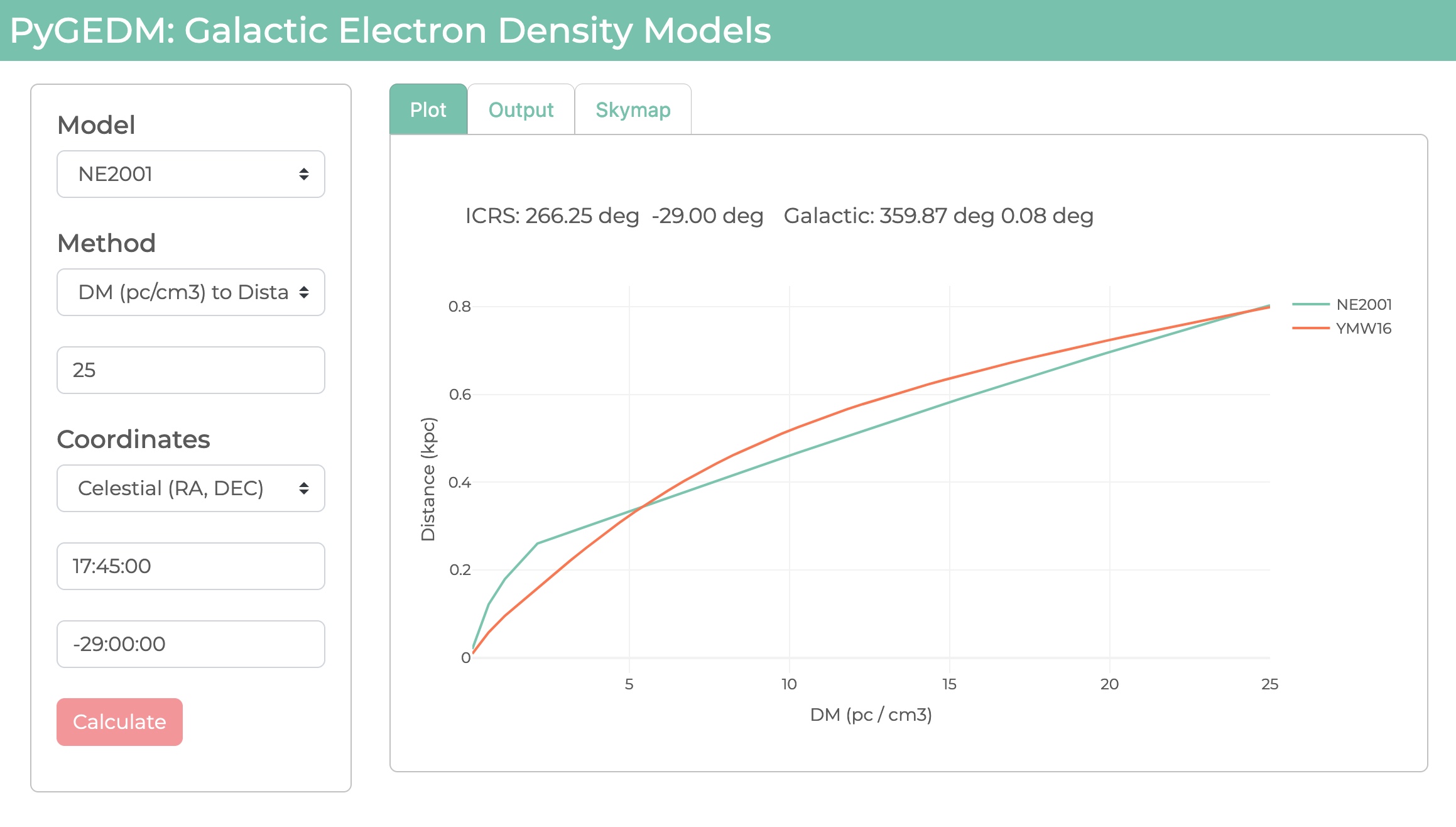}%
\caption{Screenshot of \textsc{PyGEDM} web app, with example output.}
\label{fig:webapp}
\end{figure}

We provide a standalone web application for \textsc{PyGEDM} in the {\tt /app} directory of the code repository. This application can be used for estimation of DM along a line of sight to a given distance, and vice versa. The contribution of both NE2001 and YMW16 models as a function of DM (or distance) is shown graphically along the line of sight. 

The application uses {\tt plotly}\footnote{\url{https://plotly.com}} and {\tt dash-labs}\footnote{\url{https://github.com/plotly/dash-labs}} to create the website interface, which is served using the {\tt gunicorn} web server. A {\tt Dockerfile} is provided to build and run the web application.

\section{Comparison of YMW16 and NE2001 models}

\begin{figure*}
\centering 
\includegraphics[width=2.05\columnwidth]{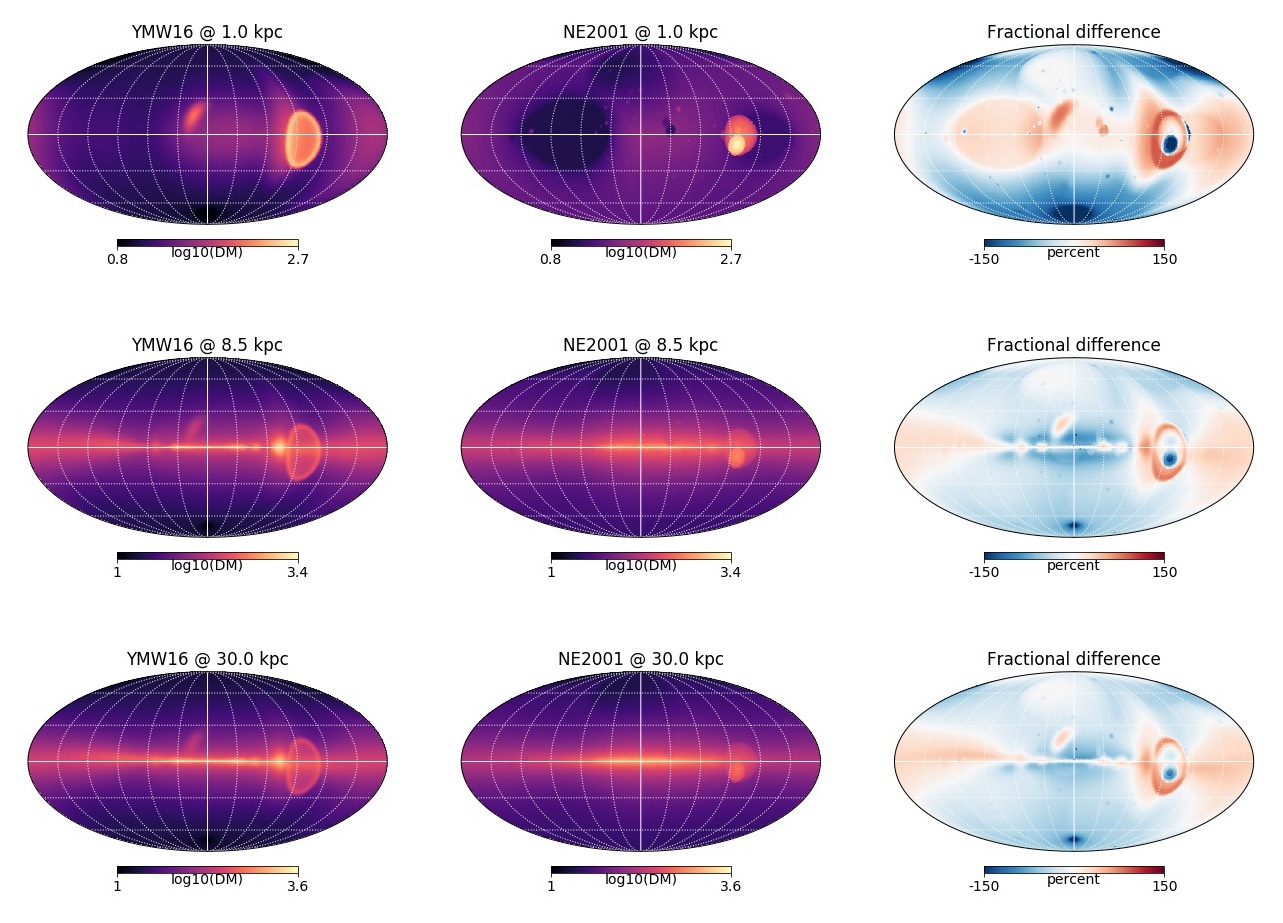}%
\caption{All-sky maps (Mollweide projection) in Galactic coordinates, showing  DM along line of sight to 1 kpc (top), 8.5 kpc (middle), and 30 kpc (bottom), for the YMW16 (left) and NE2001 (center) models. Fractional difference between the two maps is shown on the right.}
\label{fig:allsky-comparison}
\end{figure*}

\begin{figure}                                                                                                   
\centering 
\includegraphics[width=1.0\columnwidth]{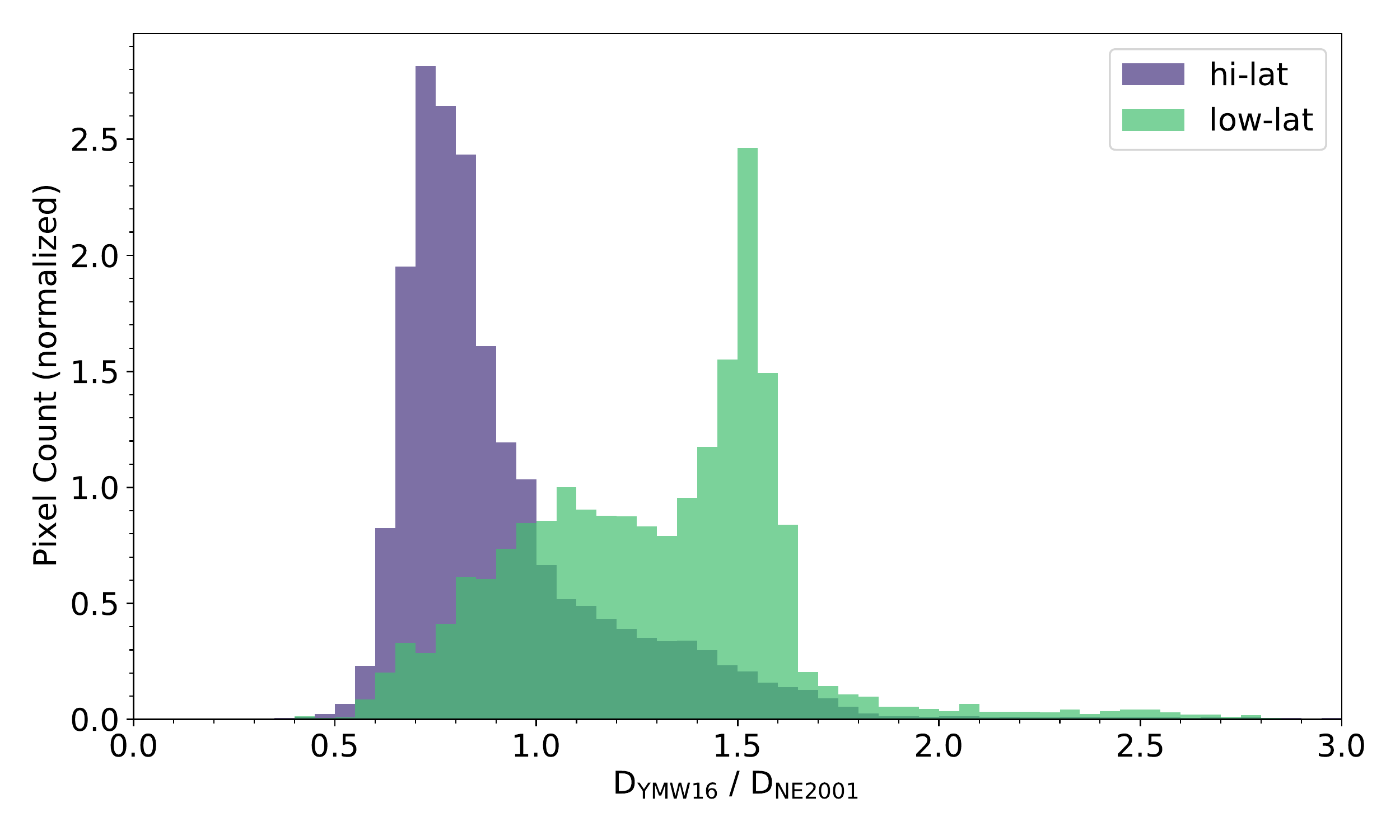}%
\caption{Histograms of $D_{\rm{YMW16}}$/$D_{\rm{NE2001}}$, the ratio of model distance prediction for the YMW16 and NE2001 models. On average, at low Galactic latitude ($|b| < 2^{\circ}$, green), YMW16 predicts larger distances than NE2001; at high latitudes ($|b| > 2^{\circ}$, purple), YMW16 predicts smaller distances. }
\label{fig:model-lat-ratios}
\end{figure}


Fig.\,\ref{fig:allsky-comparison} shows all-sky images made from YMW16 and NE2001, made using \textsc{PyGEDM} and plotted using the \textsc{Healpy} package. Averaged across all pixels, the NE2001 model returns a mean DM of 107.9\,\udm\, and median 55.3\,\udm\, when evaluated out to its 17\,kpc extent. The YMW16 model returns 110.6\,\udm\, and 48.7\,\udm, respectively, when evaluated out to 30\,kpc. That is, NE2001 reports a median $\rm{DM_{ISM}}$ contribution 1.14$\times$ higher than YMW16. There is, however, a strong dependence on Galactic latitude. At low latitudes ($|b|<2^{\circ}$), YMW16 predicts larger distances than NE2001, whereas at high latitudes YMW16 predicts smaller distances (Fig.\,\ref{fig:model-lat-ratios}).

As apparent in Fig.\,\ref{fig:allsky-comparison}, at 1\,kpc large fractional differences\footnote{The fractional difference is defined as $\rm{(DM_{YMW16} - DM_{NE2001}) / DM_{YMW16}}$; negative values imply $\rm{DM_{NE2001} > DM_{YMW16}}$.} between the two models are returned at the location of NE2001 clumps, with YMW16 in excess along most lines of sight at low Galactic latitudes, corresponding to features such as the Gum nebula, local bubble, Loop I, and NE2001's low density region.  

At 8.5\,kpc (i.e. the distance to the Galactic centre), it becomes apparent that YMW16 predicts higher DM values away from Galactic centre ($|l|<90^{\circ}$). YMW16 predicts higher DMs at low latitudes, but NE2001 predicts higher DMs in most other directions. These general trends remain out to 30\,kpc (lower panels). 

\begin{figure}
\centering 
\includegraphics[width=1.0\columnwidth]{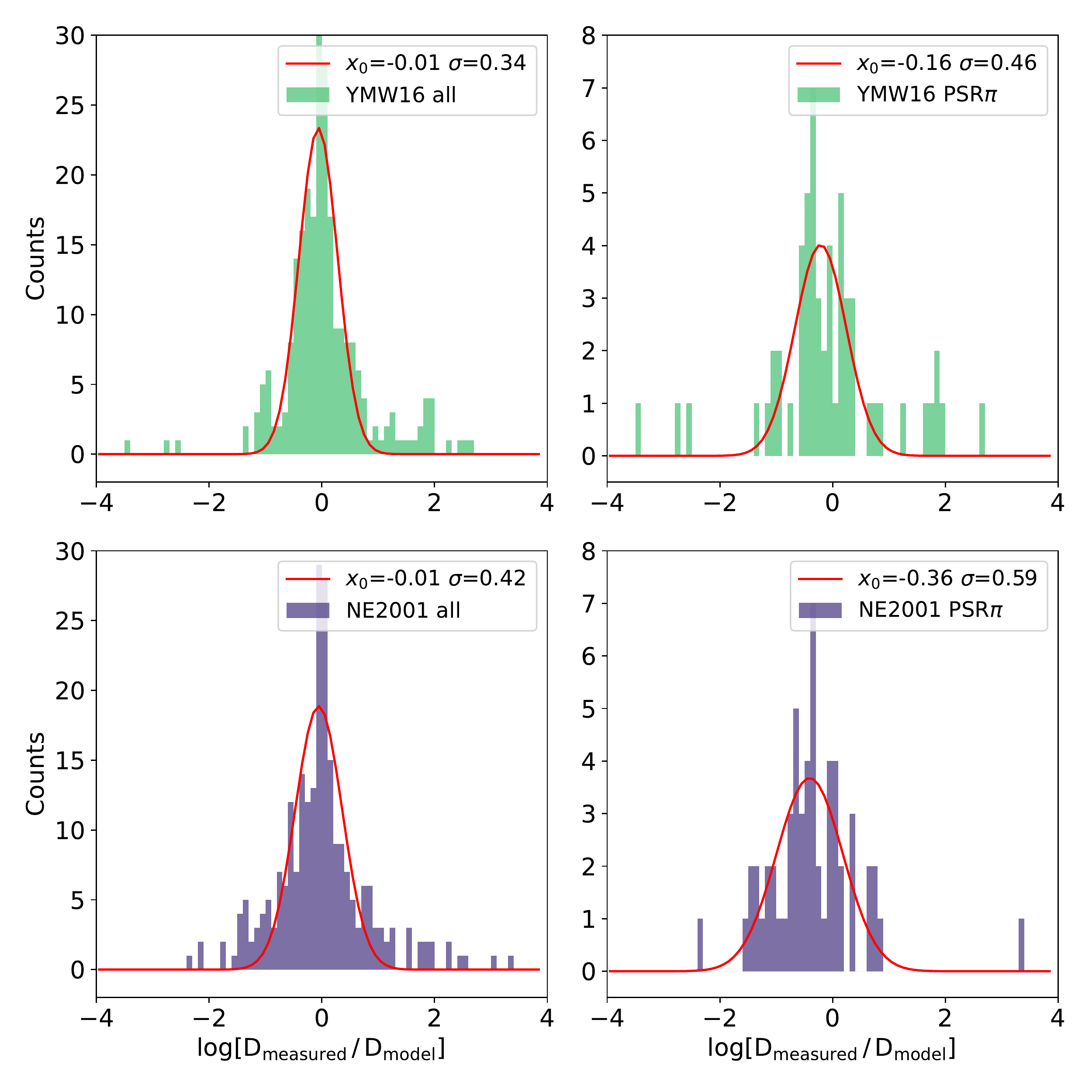}%
\caption{Histograms of log$_e$($D_{\rm{measured}}$/$D_{\rm{model}}$), the ratio of model-independent measured distance to the model estimate for the 189+57 pulsar sample (top panels) and 57 PSR$\pi$ sample (bottom panels). Gaussian fits to the histograms are shown in red.}
\label{fig:model-vs-measured}
\end{figure}

We may also compare model predictions for pulsar distance to GEDM-independent distance measurement. The PSR$\pi$ survey obtained parallax-based distance measurements for 57 pulsars \citep{Deller:2019}, which complements the 189 measurements used in YMW16. To estimate errors $\varepsilon_i$, we apply a log transform to normalise the data:
\begin{equation}
    \mathrm{log}_e(D_{\rm{measured},i}) = \mathrm{log}_e(D_{\rm{model},i}) + \varepsilon_i\label{eq:log-err}
\end{equation}
Fig.\,\ref{fig:model-vs-measured} shows a histogram of $\mathrm{log}_e(D_{\rm{measured}}$/$D_{\rm{model}})$, the ratio of measured distance to model prediction, for the 189+57 measurements (left panels), and for the PSR$\pi$ sample (right panels). Corresponding Gaussian fits are also plotted, showing that the log transform has normalised the data. From Eq.\,\ref{eq:log-err}, treating the error term as $\varepsilon_i = \mu \pm N\sigma$, where $\sigma$ and $\mu$ are the standard deviation and expected value of the Gaussian fit, and $N$ is the number of standard deviations used in confidence interval, we find:
\begin{equation}
    D_{\rm{model},i} \, e^{\mu - N\sigma} \leq D_{\rm{measured},i} \leq D_{\rm{model},i} \, e^{\mu + N\sigma}
\end{equation}
As the distribution of $\varepsilon_i$ is not perfectly Gaussian, 3$\sigma$ does not correspond to a 99\% confidence interval (CI); as such we report CI percentages. 
For pulsars within the 189+57 sample, using 3$\sigma$ we find that 86\% of YMW16 distance estimates lie between 
0.35--2.76$\times D_{\rm{measured}}$; 
for NE2001, 87\% of distance estimates lie between 
0.28--3.48$\times D_{\rm{measured}}$. 
That is, YMW16 performs better on average. 


Nonetheless, this analysis of all 189+57 pulsars is biased toward YMW16 as it includes the pulsars used in the creation of YMW16. As discussed in \citet{Deller:2019}, the PSR$\pi$ pulsars can be used as an independent test, as they were not used in the creation of either model. When considering only the PSR$\pi$ sample (Fig.\,\ref{fig:model-vs-measured}, right panels), 86\% of NE2001 distance estimates lie within 
0.12--4.10$\times D_{\rm{measured}}$, with a mean offset of $0.70$; i.e. NE2001 systematically underestimates distances. YMW16, in comparison, has a mean offset of $0.85 $; however, the Gaussian fit has several significant outliers. The 3$\sigma$ range for YMW16 is 0.22--3.33$\times D_{\rm{measured}}$, with 82\% of estimates within this range. 

The most discrepant pulsars (where the model DM estimate differs by more than an order of magnitude) for the two GEDMs are summarized in Tab.\,\ref{tab:outliers}. Of these:
\begin{itemize}
    \item J0248$+$6021 is located in/behind a nebula toward the Galactic anticenter; both models greatly overestimate its distance.
    \item J0942$-$5552 and J1017$-$7156 are located toward the Gum nebula ($l\sim264^{\circ}$). NE2001 places J0942$-$5552 significantly further away, and both models overestimate distance to J1017$-$7156.
    \item J1735$-$0724, and J1741$-$0840 are located toward Loop I.
    \item J1623$-$0908 lies behind a the HII Region Sh 2-27 \citep{KochOcker:2020}.
    \item J1745$-$3040 is located toward the Galactic center.
\end{itemize}

The two pulsars J1735$-$0724 and J1741$-$0840 are particularly poorly estimated by YMW16. The discrepancy appears to be due to excess electron density below 200\,pc, due to the contribution of Loop I. The discrepancy suggests that either the electron density of Loop I is overestimated and/or that Loop I is further away than modeled. We highlight this as of particular interest, given that the distance to Loop I is contentious  \citep[e.g.][]{Bland-Hawthorn:2003, Shchekinov:2018, Dickinson:2018}.

\begin{table}

\caption{\label{tab:outliers} Table of most significant outliers, where $D_{\rm{model}}$/$D_{\rm{measured}}$, the ratio of model prediction to measured distance, is below 0.1 or greater than 10. Bolded values indicate where one model notably better predicts the distance. Pulsars from the PSR$\pi$ sample are marked with an asterisk.
}

\begin{tabular}{lccc}
\hline
Source &  D$_{\rm{meas}}$ & D$_{\rm{NE2001}}$ &  $D_{\rm{YMW16}}$  \\
       & (pc)           & (pc)             & (pc)   \\
\hline
\hline

J0248+6021 &   $2000_{-200}^{+200}$  & 43459 &  25000  \\ 
J0942$-$5552 &    $300.0_{-200}^{+800}$ & 3770 & \textbf{415}   \\ 
J1017$-$7156 &    $256^{-60}_{+114}$  & 2980   & 1807  \\ 
J1623$-$0908* & $1710_{-250}^{+2050}$ & 49998   &  25000  \\ 
J1735$-$0724* & $6680_{-5250.0}^{+8710.0}$ & \textbf{2261}  &   213  \\ 
J1741$-$0840* & $3580_{-550}^{+940}$ &  \textbf{2169}       &  222   \\
J1745$-$3040 & $200_{-200}^{+1100}$  & 1913    & 2343 \\ 
J1912+2104* & $41020_{-35900}^{+377750}$ & 3960 &   3360  \\ 
\hline
\end{tabular}
\end{table}

\begin{figure*}
\centering 
\includegraphics[width=2.05\columnwidth]{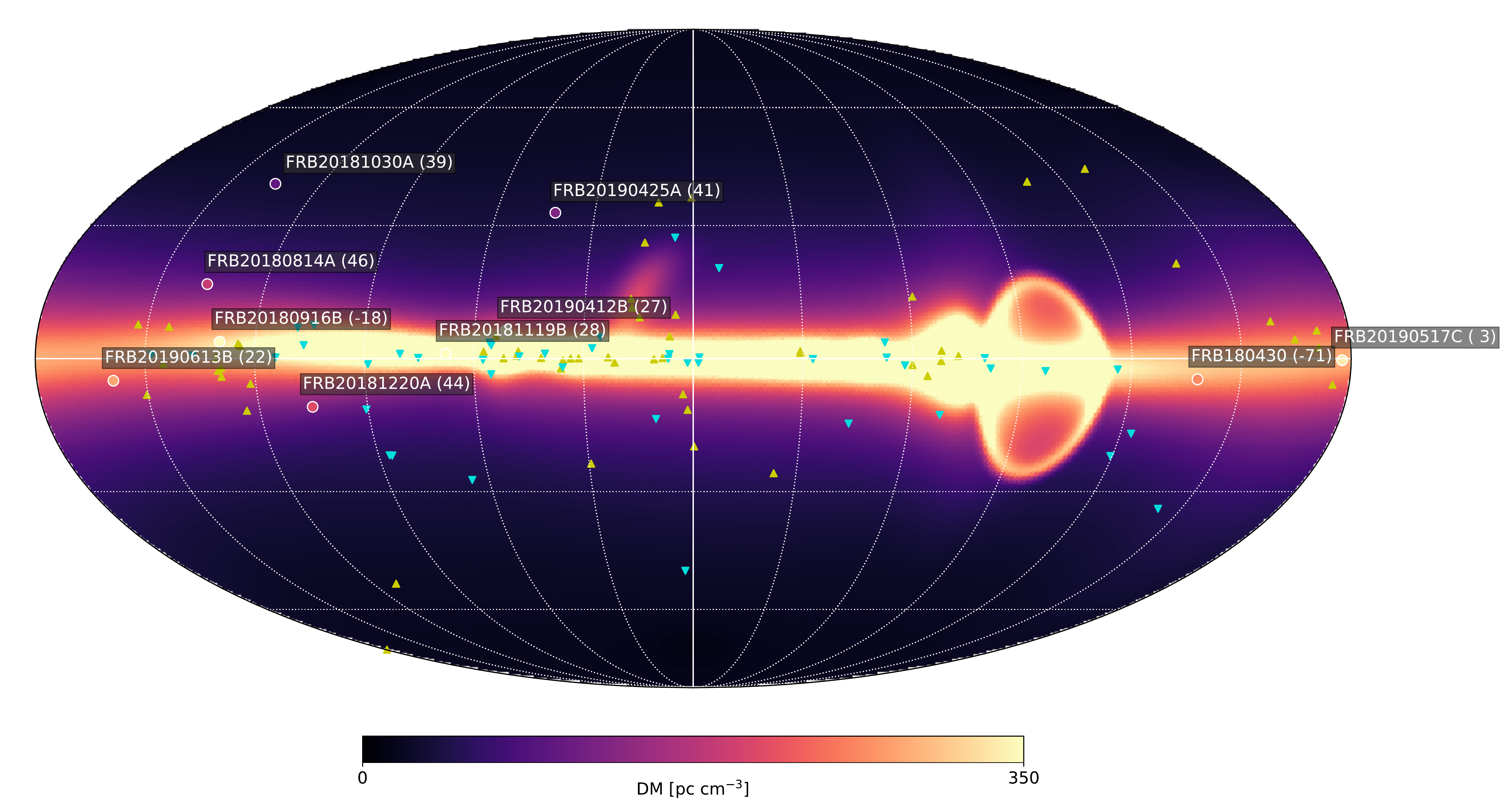}%
\caption{Location of FRBs with low DM excess ($<$50 \udm), plotted on top of total Galactic DM contribution (YMW16 + YT20). Also overlaid are pulsars where YMW16 distance is overestimated (gold $\blacktriangle$), or underestimated (cyan $\blacktriangledown$) by more than 1.5$\times$. The DM excess, in \udm, for each FRB is shown in parentheses. }
\label{fig:frb-locations}
\end{figure*}


We find some spatial clustering of pulsars for which the GEDMs poorly predicts pulsar distance (Fig.\,\ref{fig:frb-locations}). In Fig.\,\ref{fig:frb-locations},  pulsars where distances are over- or underestimated by more than 1.5$\times$ are plotted on top of a map of $\rm{DM_{MW}}$ (YMW16 with YT20 halo added). Pulsars toward the anticentre are more likely to have their distances overestimated, as are those in the direction of Loop I (for YMW16 only) and the Gum nebula. 

To look for trends, we took the subset of pulsars with distance estimates discrepant by more than 1.5$\times$ (Fig.\,\ref{fig:frb-locations}),  and binned them by latitude, longitude, and distance. We find:
\begin{itemize}
    \item Pulsars between
 $60^{\circ}<l<90^{\circ}$ are more likely to have their distances underestimated for both models.
 \item Pulsars between $120^{\circ}<l<180^{\circ}$ are more likely to have distances overestimated for YMW16. NE2001 is likely to overestimate between $120^{\circ}<l<150^{\circ}$.
 \item Low Galactic latitudes ($b < -15^{\circ}$) are likely to be underestimated by YMW16, but overestimated by NE2001. 
 \item Pulsars with distances below 1 kpc are more likely to be underestimated by both models, whereas pulsars with distances between 1--3 kpc are likely to be overestimated.
\end{itemize}

We also plot FRBs with low DM excess ($<$50 \udm) in Fig.\,\ref{fig:frb-locations}, taken from FRBCAT\footnote{\url{http://www.frbcat.org}} on 2021-06-10 \citep{Petroff:2016}, and the CHIME FRB Catalog 1 \citep{CHIME:2021}. The combined $\rm{DM_{MW}}$ map places two FRBs within the galaxy: FRB20180916B (previously named FRB180916.J0158+65) \citep{CHIME:2019} and FRB180430 \citep{Qiu:2019}. Both of these FRBs are at low latitudes on sight lines away from the Galactic centre. FRB20180916B has been shown to be extragalactic \citep{Marcote:2020}; taken together with pulsar distance overestimates, one can conclude that YMW16 systematically overestimates on sight lines toward the Galactic anticentre. In contrast, NE2001 does not place any FRB within the Galaxy.

\section{Discussion}

Both NE2001 and YMW16 have proven to be invaluable tools for conversion between DM values and distance estimates. In this article, we introduce \textsc{PyGEDM}, a Python package that provides a unified interface to NE2001, YMW16, and the YT20 GEDMs. 

We used \textsc{PyGEDM} to quantitatively compare and contrast the NE2001 and YMW16 models, highlighting where their predictions differ. We also compare GEDM predictions to pulsar measurements, finding that YMW16 performs better on average than NE2001, but that both models show significant outliers. We highlight that distances to J1735$-$0724 and J1741$-$0840 are poorly estimated by YMW16, suggesting that Loop I may be further away than modelled. There is still debate as to whether Loop I is a local feature--as YMW16 assumes---or associated with the Fermi bubble above Galactic centre, or both \citep{Dickinson:2018, Shchekinov:2018}. Additionally, pulsars between $120^{\circ}<l<180^{\circ}$ are more likely to have distances overestimated for YMW16. Taken together with FRB measurements, it is clear that YMW16 overestimates the overall Galactic DM contribution toward the Galactic anticentre.

Both NE2001 and YMW16 underestimate distance for pulsars in the PSR$\pi$ sample.  As discussed in \citet{Deller:2019}, pulsars at high galactic latitudes are over-represented in $PSR\pi$, so it is not an unbiased sample. The median distance for pulsars in the sample is 2.5\,kpc, whereas the median distance for pulsars with parallax measurement used in YMW16 is 1.1\,kpc. Incorporating the PSR$\pi$ sample will improve next-generation GEDMs, particularly at high latitude.

More generally, new GEDM-independent pulsar distance measurements will provide tests of GEDMs and further data to use in modelling. We suggest that pulsar targets are chosen strategically, focusing on areas where the GEDMs give poor distance estimates; namely, toward the Galactic anticentre, Loop I, the Gum nebula, and at low Galactic latitudes.

Improving models of DM$_{\rm{halo}}$ will be important for FRB-based cosmology experiments. \citet{Kumar:2019} argue that biases in any noncosmological contributions to DM must be kept to under 0.6\%: a budget of only 15.6\,\udm\,for even the highest recorded DM of $2596.1\pm0.3$\,\udm\, \citep[FRB160102,][]{Bhandari:2018}. As such, future GEDMs should also ensure that their estimate of DM$_{\rm{ISM}}$ does not contain contributions from the Galactic halo; or, a combined ISM and halo parameter could be fit to data. Currently, simply adding the YT20 to YMW16/NE2001 DM estimate likely overestimates the Galactic contribution for a given FRB.

Current GEDMs do not provide the user the tools to rerun parameter fitting with additional data or to add/modify features. We suggest that future GEDMs should provide such tools, so new measurements can be rapidly incorporated to improve the model. 

Follow-up campaigns toward repeating FRBs will result in long observations, upon which pulsar searches could be conducted. Searches for pulsars along or near FRB sight lines would allow more accurate determination of $\rm{DM_{ISM}}$ toward the FRB. We suggest that such searches should be done as a matter of course, to improve our understanding of the ISM and help facilitate the emerging field of FRB cosmology.

\begin{acknowledgements}
We thank the authors of NE2001 (J.Cordes and J. Lazio), YMW16 (J. M. Yao, R. N. Manchester, N. Wang), YT20 (S. Yamasaki and T. Totani) and PZ19 (J. X. Prochaska and Y. Zheng) for email correspondence and providing these valuable models. A.T.D. is the recipient of an Australian Research Council Future Fellowship (FT150100415).
\end{acknowledgements}

\bibliographystyle{pasa-mnras}
\bibliography{references}

\end{document}